\documentclass[12pt]{iopart}
\usepackage{iopams}
\expandafter\let\csname equation*\endcsname\relax
\expandafter\let\csname endequation*\endcsname\relax 
\usepackage{graphicx,amsmath,braket,bbm,easy-todo,svg} 

\graphicspath{{Figures/}}

\usepackage{xcolor}
\usepackage{hyperref}
\hypersetup{
  colorlinks   = true, 
  urlcolor     = blue, 
  linkcolor    = blue, 
  citecolor   = blue 
}

\usepackage[square, numbers, comma, sort&compress]{natbib}

\let\OLDthebibliography\thebibliography
\renewcommand\thebibliography[1]{
  \OLDthebibliography{#1}
  \setlength{\parskip}{0pt}
  \setlength{\itemsep}{0pt plus 0.5ex}
}

\DeclareMathOperator{\sign}{sign}

\begin{document}

\title{Operator dynamics and entanglement in space-time dual Hadamard lattices}

\author{Pieter W. Claeys}
\address{Max Planck Institute for the Physics of Complex Systems, 01187 Dresden, Germany}
\ead{claeys@pks.mpg.de}

\author{Austen Lamacraft}
\address{TCM, Cavendish Laboratory, University of Cambridge, Cambridge CB3 0HE, UK}
\ead{al200@cam.ac.uk}

\begin{abstract}
Many-body quantum dynamics defined on a spatial lattice and in discrete time -- either as stroboscopic Floquet systems or quantum circuits -- has been an active area of research for several years. Being discrete in space and time, a natural question arises: when can such a model be viewed as evolving unitarily in space as well as in time? Models with this property, which sometimes goes by the name space-time duality, have been shown to have a number of interesting features related to entanglement growth and correlations. One natural way in which the property arises in the context of (brickwork) quantum circuits is by choosing dual unitary gates: two site operators that are unitary in both the space and time directions.

We introduce a class of models with $q$ states per site, defined on the square lattice by a complex partition function and paremeterized in terms of $q\times q$ Hadamard matrices, that have the property of space-time duality. These may interpreted as particular dual unitary circuits or stroboscopically evolving systems, and generalize the well studied self-dual kicked Ising model. We explore the operator dynamics in the case of Clifford circuits, making connections to Clifford cellular automata \cite{schlingemann2008structure} and in the $q\to\infty$ limit to the classical spatiotemporal cat model of many body chaos \cite{gutkin2021linear}. We establish integrability and the corresponding conserved charges for a large subfamily and show how the long-range entanglement protocol discussed in the recent paper \cite{lotkov2022floquet} can be reinterpreted in purely graphical terms and directly applied here.
\end{abstract}

\section{Introduction}

The existence of symmetries between continuous space and time in fundamental physical theories has been with us for more than a century, since the dawn of relativity theory. In this work we are concerned with models –– either in classical or quantum mechanics –– defined in \emph{discrete} space and time where a similar symmetry arises. 

Lattice models studied in statistical mechanics usually have a symmetry between different lattice directions. When studying \emph{dynamics} of lattice systems there is typically an inherent asymmetry between the discrete lattice directions and the continuous time direction. A more symmetric situation arises for discrete time dynamics \footnote{The term \emph{stroboscopic} is often used when the discrete time dynamics arises from continuous time evolution in a non-autonomous system}. Taking the case of one space ($x$) and one time ($t$) dimension as our main example, with $x, t\in \mathbb{Z}$, there is no guarantee in general that the dynamical map that advances the system forward one unit of time $t\to t+1$ can be reformulated as a map advancing the system forward in space $x\to x+1$.  In this paper we will be concerned with particular families of systems where such a reformulation is possible, a situation that we will describe as \emph{space-time duality}.

Two questions immediately arise:

\begin{enumerate}
\item How is this reformulation defined, particularly in the quantum mechanical case?
\item What, apart from intrinsic interest, is the value in such a reformulation?
\end{enumerate}

The answer to the first question is deferred until \sref{sec:model}, where the class of models that we will study is defined precisely. As to the second question, systems that may be given such a space-time dual formulation have particular dynamical features, particularly in regard to the propagation of correlations and (in the quantum case) the growth of entanglement between distinct regions. Further, these features mean that the calculation of these quantities is dramatically simpler than for generic many-body systems, even though these systems do not fall into one of the usual categories of systems (e.g. free particles, integrable models, etc.) where such calculations are often possible, though they may intersect with them in some special cases.

Space-time duality in the sense used in this paper made early appearances in \cite{hosur2016chaos,gutkin2016classical} in the quantum and classical settings respectively. Its occurrence in the kicked Ising model at the self-dual point was uncovered in \cite{akila2016particle} and was subsequently used in \cite{bertini2018exact} to calculate the spectral form factor and in \cite{bertini2019entanglement} to evaluate the (maximal) growth of entanglement. Ref.~\cite{gopalakrishnan2019unitary} noticed that in the brickwork circuit formulation of this system space-time duality arose from a ``dual unitary'' condition on the unitary gates describing the circuit, which was then used to study correlations in a much wider range of models in \cite{bertini2019exact}. For the kicked Ising model this space-time duality proved to be crucial in deriving analytic results on deep thermalization~\cite{ho_exact_2022} and measurement-based quantum computing~\cite{stephen2022universal}. A more extensive list of references describing subsequent activity introducing new families of dual unitary circuits and calculating their properties can be found in Section 3.3.3 of the recent review \cite{fisher2023random}.

The models discussed in this paper have their origin in \cite{gutkin2020exact}, which showed that models based on Hadamard matrices (as described in \sref{sec:model}) have space-time duality. Ref.~\cite{aravinda2021dual} introduced the quantum cat maps as an ingredient in constructing dual unitaries and sucessive generalizations were explored in \cite{borsi2022construction,claeys2022emergent,claeys2023dual}. Quantum circuits in terms of Hadamard matrices also allow for extending dual unitarity to e.g. different lattices and higher dimensions~\cite{liu2023solvable,osipov_correlations_2023,rampp_entanglement_2023,sommers_zero-temperature_2024}, closely connected to recent `hierarchical' generalizations of dual-unitarity~\cite{yu_hierarchical_2024}. 

Here, we focus on the operator dynamics and entanglement generation properties of such models for the following reasons:

\begin{itemize}
\item It is possible to handle different local Hilbert space dimensions $q$ on a unified footing, even including the classical limit $q\to\infty$ where we can make a connection to the spatiotemporal cat models introduced in \cite{gutkin2016classical} (see also \cite{gutkin2021linear}).

\item These models include families of Clifford space-time dual models for arbitrary $q$, generalizing the discussion given in \cite{gutschow2010time,sommers2023crystalline} for $q=2$. These models have simple operator dynamics (without operator entanglement) but include a variety of behaviours such as the ballistic propagation of operators or fractal operator growth.

\item Certain members of the family are integrable, in the sense that they allow for the construction of an extensive set of local conserved charges. These models include known integrable limits of the kicked Ising model ($q=2$) and the kicked Potts model ($q=3$). More generally we show how each complex Hadamard matrix can be used to construct an integrable model, extending these constructions to arbitrary $q$ and general complex Hadamard matrices (in a way that will be made more precise in \sref{sec:lotkov}).

\item In these models the preparation of highly entangled states under certain protocols can be understood using purely graphical methods, which are however different from those used to date to understand entanglement growth in dual unitary circuits (see e.g. \cite{bertini2019entanglement,gopalakrishnan2019unitary}).

\end{itemize}

\subsection{Outline}

The outline of the remainder of this paper is as follows. In the next section we introduce the models that we will study,  review some of the properties of the Hadamard matrices used to define them, and give some examples. Section~\ref{sec:pauli} introduces the generalized Pauli operators as a convenient operator basis for a $q$-dimensional local Hilbert space and investigates Hadamard matrices that act in a simple way on this basis, including Clifford and quantum cat maps. In \sref{sec:clifford} we study the operator dynamics in circuits constructed from these ingredients, finding a variety of behaviours and making connections to Clifford cellular automata \cite{gutschow2010time,sommers2023crystalline} and the spatiotemporal cat maps of \cite{gutkin2016classical,gutkin2021linear} in the classical limit.  
In \sref{sec:lotkov} we consider unitary dynamics obtained from a particular arrangement of Hadamard matrices and show how these support solitonic dynamics, giving rise to an extensive set of conserved charges. We discuss the Floquet dynamics of a $q=3$ model introduced in \cite{lotkov2022floquet} that turns out to be an example of this construction, which leads to a simple graphical derivation of the results of that paper on long-ranged entanglement generation. In this way this protocol directly extends to all models constructed in that section.
We conclude in section~\ref{sec:conclude} and offer some perspectives for future work.

\section{Model} \label{sec:model}

In this work we will be concerned with models defined on the square lattice that may be interpreted as discrete time unitary evolution in 1+1 dimensions with a finite local Hilbert space of dimension $q$. The general form of the (complex) partition function of these models is
\begin{equation}
U = \mathcal{N}\sum_{z_i\in \mathbb{Z}_q}\prod_{\braket{i,j}} u_{ij}(z_i, z_j),
\label{eq:partition}
\end{equation}
where $\mathcal{N}$ is a normalization factor. As usual $\braket{i,j}$ denotes the bond between site $i$ and $j$ of the lattice, where the site $i= (x,t)$ is specified by two integer coordinates. The function $u_{ij}(z_i, z_j): \mathbb{Z}_q\times \mathbb{Z}_q\longrightarrow U(1)$ is a phase factor describing the bond $\braket{i,j}$. We will mostly be concerned with the case when all space-like (horizontal) bonds are described by the same function $u_\text{H}$ and likewise all time-like (vertical) bonds are described by a function $u_\text{V}$. 

The quantum mechanical interpretation of \eref{eq:partition} is as follows. We associate with each lattice site the Hilbert space $\mathbb{C}^q$ (a `qudit') with an orthonormal basis $\ket{z}$, $z\in\mathbb{Z}_q$ (the \emph{computational basis}). The phase factors $u_{ij}(z_i,z_j)$ are regarded as the matrix elements of an operator $u_{ij}$ in the computational basis: $u_{ij}(z_i,z_j)=\braket{z_i|u_{ij}|z_j}$. $N$ qudits in a row of the lattice are described by a Hilbert space spanned by the tensor product states $\ket{z_{1:N}}\equiv\ket{z_1}\otimes\ket{z_2}\cdots \otimes \ket{z_N}$.

\begin{figure}
  \begin{center}
  \includegraphics[width=0.65\textwidth]{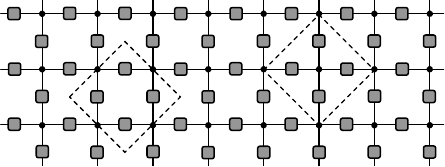}
  \caption{The partition function \eref{eq:partition} represented as a tensor network, with rank 2 tensors on the bonds given by the operators $u_{ij}$ denoted by boxes and rank 4 tensors $\delta_{z_1,z_2,z_3,z_4}$ at the vertices denoted by dots. The network can be viewed as a `brickwork' circuit by taking the elementary `unit cell' to be the dotted square on the left (\sref{sec:brickwork}). 
  Alternatively, the dotted square on the right shows a decomposition into  `round-a-face' form (see \sref{sec:clockwork}). 
  \label{fig:tn}}
  \end{center}
\end{figure}

The partition function in \eref{eq:partition} is equal to the contraction of a tensor network with rank 2 tensors on the bonds given by the operators $u_{ij}$ and rank 4 tensors $\delta_{z_1,z_2,z_3,z_4}$ at the vertices, as illustrated in Fig.~\ref{fig:tn}. The $p$-valent $\delta$ tensor is defined as
\begin{equation}
  \delta_{z_1,z_2,\ldots ,z_p} = 
  \begin{cases}
    1  & z_1=z_2=\cdots =z_p \\
    0  & \text{otherwise}
  \end{cases}
\end{equation}
In \eref{eq:partition} we did not specify the boundary conditions. We fix the values of the $z_{x,t}$ on the bottom and top row to $z_{x,0}$ and $z_{x,T}$ respectively for $x=1,\ldots N$, thus omitting them from the sum in Eq.~\eqref{eq:partition}. We take periodic boundary conditions in the spatial direction so that $z_{x,t}=z_{x+N,t}$. We want the resulting partition function $U(z_{1:N,0},z_{1:N,T})$ to give the matrix elements of a unitary operator on a system of $N$ qudits. This will be the case if certain conditions on the $u_{ij}$ are satisfied. A single row in the tensor network corresponds to a diagonal operator $U_{\text{row}}$
\begin{align}
  U_{\text{row}}  = \vcenter{\hbox{\includegraphics[width=0.65\linewidth]{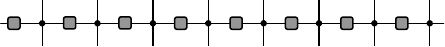}}}\, ,
  \end{align}
with matrix elements
\begin{equation}
  \braket{z_{1:N}|U_{\text{row}}|z_{1:N,t}} = \prod_{x=1}^N u_\text{H}(z_{x,t},z_{x+1,t})
\end{equation}
This is a unitary operator since the functions $u_\text{H}$ are phases. The vertical bonds correspond to operators $u_\text{V}$ acting on single qudits, giving rise to an operator $U_\text{vert}$ acting on $N$ qudits
\begin{align}
  U_{\text{vert}}  = \quad \vcenter{\hbox{\includegraphics[width=0.62\linewidth]{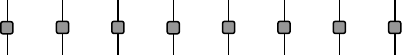}}}\, .
  \end{align}
To obtain unitary evolution we require that $u_\text{V}$ is unitary too, up to a multiplicative factor. A $q$-dimensional matrix with elements of unit modulus obeying $u_\text{V}^\dagger u_\text{V}=q\mathbbm{1}$ is called a \emph{complex Hadamard matrix}. A simple and important example is the Fourier matrix
\begin{equation}
\left(F_q\right)_{jk} = \exp\left(2\pi ijk/q\right)\qquad j,k=0,\ldots, q-1.
\end{equation}
Acting with $F_q$ on a $q$-dimensional vector performs the discrete Fourier transform (or its inverse in the usual convention). 

The complete tensor network then corresponds to the composition 
\begin{equation}\label{eq:big-U}
U(z_{1:N,0},z_{1:N,T})\propto\braket{z_{1:N,T}|\left(U_\text{vert}U_\text{row}\right)^T|z_{1:N,0}}.
\end{equation}
These are the matrix elements of a unitary operator if $u_V$ is a Hadamard matrix, with the normalization factor in \eref{eq:partition} given by $\mathcal{N}=q^{-N(T-1)}$.

We may consider evaluating the tensor network from left to right instead of from bottom to top. Without worrying about boundary conditions at this stage, we can say that this space-like evolution is unitary if $u_\text{H}$ is Hadamard. Thus if both $u_\text{V}$ and $u_\text{H}$ are Hadamard the model describes unitary evolution in both space and time. This property -- known as \emph{space-time duality} -- has been exploited in various contexts in recent years, as described in the introduction. For introductions to complex Hadamard matrices, we refer the reader to~\cite{tadej2006concise,banica_invitation_2023}.

\subsection{Interpretation as Stroboscopic Dynamics}

We have introduced our model abstractly, but an interpretation in terms of stroboscopic dynamics is straightforward if we can write $U_\text{row}$ and $U_\text{ver}$ as the exponentials of two Hamiltonians
\begin{align}\label{eq:strob}
  U_{\text{row}} = e^{-i H_1 T_1}, \qquad  U_{\text{vert}} = e^{-i H_2 T_2}\,.
  \end{align}
As discussed above, the partition function \eref{eq:partition} describes space-time dual unitary evolution if $U_{\text{vert}}$ is a product of single qudit Hadamards and $U_{\text{row}}$ is diagonal with matrix elements a product of two qudit phase functions $u_\text{H}(z_{x}, z_{x+1})$, both with respect to the computational basis. Thus $H_1$ is a sum of diagonal terms acting on two neighbouring qudits, while $H_2$ is a sum of single qudit terms. The former can be interpreted as a classical Ising-like interaction, with the latter acting as a `kick' creating a quantum superposition. Further, $H_{1,2}$ must be chosen so that the matrix elements of the corresponding unitaries have unit modulus. Concrete examples include the self-dual kicked Ising model \cite{akila2016particle} for $q=2$ and the kicked Potts model studied in \cite{lotkov2022floquet} and \sref{sec:lotkov} for $q=3$. Space-time duality therefore requires a degree of fine tuning, including the parameters $T_1$ and $T_2$ in \eref{eq:strob}.

\subsection{Interpretation as a Brickwork Circuit}\label{sec:brickwork}

Alternatively, our model may be interpreted as a dual-unitary brickwork circuit. The local two-site gates constituting this circuit are defined as
\begin{align}
  \vcenter{\hbox{\includegraphics[width=0.31\linewidth]{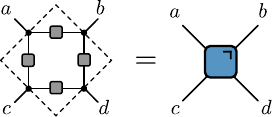}}},
\end{align}
where the vertices are now rank 3 delta tensors \cite{claeys2022emergent,claeys2023dual}. This gate is repeated in a brickwork pattern in the square lattice of Fig.~\ref{fig:tn}. The dual-unitary of the individual gate implies that this gate acts as a unitary along both the horizontal and vertical direction, as follows directly from the Hadamard construction.

The boundary conditions (e.g. initial and final states) of the two presentations of the model are however slightly different: when presented as a brickwork circuit every other horizontal Hadamard is missing from the top and bottom edges. Consider for example the two equivalent circuits below:
\begin{align}
  \vcenter{\hbox{\includegraphics[width=0.28\linewidth]{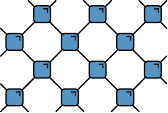}}} \qquad \Rightarrow \qquad \vcenter{\hbox{\includegraphics[width=0.28\linewidth]{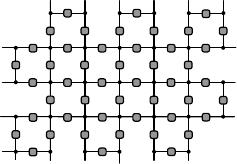}}},
\end{align}
where we take both to have periodic boundary conditions. The different boundary conditions at the initial and final time however do not change the physics of the brickwork circuit and the lattice model of Fig.~\ref{fig:tn}, since they simply correspond to a unitary transformation with the appropriate horizontal bonds.

\subsection{Round-a-face Form} \label{sec:clockwork}

Alternatively, our model can be presented in dual-unitary `round-a-face' form, in which the fundamental unit is a unitary acting on a single qudit controlled by its two neighbours~\cite{prosen2021many,claeys2023dual}, here of the form
\begin{align}\label{eq:clockworkgate}
  \vcenter{\hbox{\includegraphics[width=0.31\linewidth]{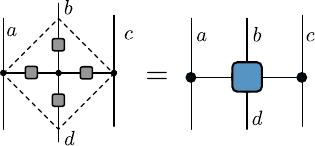}}}.
\end{align}
Ddual-unitary here implies that it is possible to exchange the role of the control indices $a,c$ and the indices of the unitary $b,d$ while retaining unitarity of the single-qudit operator -- which is again apparent in the Hadamard construction.
The full partition function of Fig.~\ref{fig:tn} reduces to regular lattice of dual-unitary interactions round-a-face with modified boundary conditions, this time of the form 
\begin{align}
  \vcenter{\hbox{\includegraphics[width=0.28\linewidth]{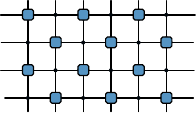}}} \qquad \Rightarrow \qquad \vcenter{\hbox{\includegraphics[width=0.28\linewidth]{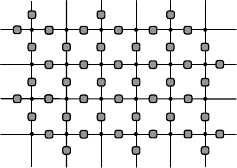}}},
\end{align}
A unified formalism that includes brickwork and round-a-face models (not necessarily constructed from Hadamard matrices) can be found in \cite{claeys2023dual}.

When the Hadamard matrices in the parametrization~\eqref{eq:clockworkgate} are chosen to be either the Fourier matrix $F$ or $F^{\dagger}$, with the four Hadamard matrices not necessarily equal, this construction gives a particularly simple family of cellular automata models. This may be seen by explicitly performing the summation over the central index in \eref{eq:clockworkgate}, where choosing e.g. all Hadamard matrices to be $F$ returns
\begin{equation}\label{eq:F-sum}
\sum_{e=0}^{q-1} \left(F_d\right)_{ae} \left(F_d\right)_{be}\left(F_d\right)_{ce}\left(F_d\right)_{de} = q\, \delta_{a+b+c+d \mod q},
\end{equation}
fixing $d = -a-b-c \mod q$. Similarly, choosing the two horizontal Hadamard matrices as $F^{\dagger}$ and the two vertical ones as $F$ returns
\begin{align}
\sum_{e=0}^{q-1} (F^{\dagger}_d)_{ae} \left(F_d\right)_{be}(F_d^{\dagger})_{ce}\left(F_d\right)_{de} = q\, \delta_{-a+b-c+d \mod q},
\end{align}
fixing $d=a-b+c$. While these two constructions are equivalent for a local two-dimensional Hilbert space, they differ in general.

To draw the correspondence to a cellular automaton described by a space-time state function $s_{x,t}\in \mathbb{Z}_q$ we identify
\begin{align}
  s_{x-1,t} = a \qquad s_{x,t+1} = b \nonumber\\
  s_{x+1,t} = c \qquad s_{x,t-1} = d
\end{align}
The relation \eref{eq:F-sum} then defines a simple `second-order'  cellular automaton (an example of a construction originally due to Fredkin \cite{margolus1984physics,vichniac1984simulating})
\begin{equation} \label{eq:second-order}
s_{x,t+1} = - s_{x-1,t} - s_{x+1, t} + s_{x,t-1} \qquad \mod q\,.
\end{equation}
For $q=2$ this corresponds to the reversible version of Wolfram's Rule 150 (Rule 150R) \cite{wolfram2002new}.

Note that this mapping involves only the variables of one sublattice of the original square lattice, with the summation being performed over indices on the other sublattice. As a result, the cellular automaton dynamics describes the evolution of particular initial product states. We will defer further discussion of this point to \sref{sec:clifford}. We now turn to more general choices of Hadamard matrices.

\subsection{Equivalent Hadamard matrices}

Before discussing possible choices of the couplings $u_\text{H}$ and $u_\text{V}$, it is useful to introduce the notion of \emph{equivalent} Hadamard matrices. $H$ and $H'$ are equivalent if
\begin{equation}\label{eq:equiv}
H' = D_1P_1 H P_2 D_2
\end{equation}
where $D_{1,2}$ are diagonal unitary matrices and $P_{1,2}$ are permutation matrices \cite{tadej2006concise}. If the same holds with $D_1=D_2=\mathbbm{1}$ we say that $H$ and $H'$ are \emph{permutation equivalent}.

A related notion is that of the \emph{dephased form} of a Hadamard matrix, in which the entries of the first row and column are all 1. If two Hadamard matrices have the same dephased form, they are equivalent. The dephased form is obtained by choosing $P_1=P_2=\mathbbm{1}$, $D_1= \operatorname{diag}(\bar H_{11},\bar H_{21},\ldots \bar H_{q1})$, and $D_2= \operatorname{diag}(1,H_{11}\bar H_{12},\ldots H_{11}\bar H_{1q})$ in \eref{eq:equiv}.

We make the following important observation about our models. It is possible for permutations to cancel at the vertices since $\delta_{z_1, z_2, z_3 \dots} = \delta_{P(z_1), P(z_2), P(z_3) \dots}$ and it is similarly possible to choose vertical and unitary Hadamard matrices such that phases also cancel. If we choose e.g. $u_{\textrm{H}} = D P u_{\textrm{H}}' P D$ and  $u_{\textrm{V}} = \bar{D} P u_{\textrm{v}}' P \bar{D}$ then all permutations and phases cancel exactly and the lattice constructed out of $u_{\textrm{H}}$ and $u_{\textrm{V}}$ is equivalent to the lattice constructed out of $u_{\textrm{H}}'$ and $u_{\textrm{V}}'$, up to boundary conditions. In this case we can always use the dephased form of the Hadamard. Such a cancellation can happen if $u_\text{V}=\bar u_\text{H}$, a case that will be discussed in detail in \sref{sec:lotkov}.

\subsection{Examples}

For $q=2,3$ and $5$ all complex Hadamard matrices are equivalent to the Fourier matrix (\cite{tadej2006concise}, section 5). The simplest case is $q=2$, where
\begin{equation}
  F_2 = \begin{pmatrix}
1 & 1 \\
1 & -1
\end{pmatrix}
\end{equation}
is the usual Hadamard gate of quantum information theory. This is equivalent to the matrix
\begin{equation}
  K_{2} \equiv \begin{pmatrix}
  1 & i \\
  i & 1
  \end{pmatrix}
  \end{equation}
since
\begin{equation}
  K_{2} = \begin{pmatrix}
  1 & 0\\
  0 & i
  \end{pmatrix} F_2\begin{pmatrix}
  1 & 0\\
  0 & i
  \end{pmatrix}.
\end{equation}
Written as a phase function this is
\begin{equation}
  K_{2}(z_i, z_j) = e^{i\pi/4}\exp\left(-\frac{i\pi}{4} \sigma_i \sigma_j\right),\qquad \sigma_i \equiv (-1)^{z_i} \in \{1, -1\}
\end{equation}
This has the form of an Ising coupling i.e. $\propto \exp\left(iJ \sigma_i \sigma_j\right)$ with a special value $J=-\pi/4$, and corresponds to the self-dual Ising model \cite{akila2016particle}.

The phase function corresponding to $F_2$ is
\begin{equation}
  F_{2}(z_i, z_j) = e^{i\pi/4}\exp\left(\frac{i\pi}{4} \left[\sigma_i \sigma_j-\sigma_i-\sigma_j\right]\right),
\end{equation}
that is, to an Ising model with a particular value of the magnetic field. The value of the magnetic field can more generally be tuned by multiplying this Hadamard matrix with diagonal unitaries.

For $q=3$ we have
\begin{equation}
  F_3 = \begin{pmatrix}
    1 & 1 & 1 \\
    1 & \omega_3 & \omega_3^2 \\
    1 & \omega_3^2 & \omega_3
  \end{pmatrix},
\end{equation}
where $\omega_q = \exp(2\pi i/q)$ is the $q$th primitive root of unity. This is equivalent to 
\begin{equation}
 K_3 = \begin{pmatrix}
  1 & \omega_3 & \omega_3 \\
  \omega_3 & 1 & \omega_3 \\
  \omega_3 & \omega_3 & 1 \\
 \end{pmatrix}
\end{equation}
Since
\begin{equation}
  F_3 = \begin{pmatrix}
    1 & 0 & 0 \\
    0 & \omega_3^2 & 0 \\
    0 & 0 & \omega_3^2
  \end{pmatrix}
K_3
\begin{pmatrix}
  1 & 0 & 0 \\
  0 & \omega_3^2 & 0 \\
  0 & 0 & \omega_3^2
\end{pmatrix}
\end{equation}
The tensor product of Hadamard matrices is a Hadamard matrix, so tensoring provides a way to build up more complicated models. The simplest example is for $q=4$
\begin{equation}
F_2\otimes F_2 = \begin{pmatrix}
1 & 1 & 1 & 1 \\
1 & -1 & 1 & -1 \\
1 & 1 & -1 & -1 \\
1 & -1 & -1 & 1
\end{pmatrix}.
\end{equation}
This differs from $F_4$ and both are dephased, however they are permutation inequivalent \cite{tadej2006permutation}. A full orbit of inequivalent Hadamards is given by \cite{tadej2006concise}
\begin{equation}\label{eq:F4_a}
  F_4^{(1)}(a)=\left(\begin{array}{cccc}
    1 & 1 & 1 & 1 \\
    1 & i  e^{i a} & -1 & -i  e^{i a} \\
    1 & -1 & 1 & -1 \\
    1 & -i  e^{i a} & -1 & i  e^{i a}
    \end{array}\right)\qquad a \in [0,\pi)
\end{equation}
$F_4^{(1)}(0)=F_4$ and $F_4^{(1)}(\pm\pi/4)$ is permutation equivalent to $F_2\otimes F_2$.

For $q=6$ the manifold of complex Hadamard matrices is quite complicated and no exhaustive classification has been found. Various constructions exist however: Complex Hadamard matrices are a specific kind of biunitary connection, and different biunitary connections can be composed to relate complex Hadamard matrices to e.g. quantum Latin squares or unitary error bases~
\cite{reutter_biunitary_2019}. There is active effort in finding new constructions which have additional properties including multi-unitarity and being perfect tensors~\cite{goyeneche_absolutely_2015,rather_construction_2023,bruzda_two-unitary_2024,bruzda2023multi}. 
\section{Generalized Pauli matrices} \label{sec:pauli}

To discuss operator dynamics with local Hilbert space dimension $q$ it is useful to introduce a basis of \emph{generalized Pauli operators}. We introduce two unitary operators $X_q$ and $Z_q$. Taking the computational basis to be the eigenbasis of $Z_q$, the two matrices have the form
\begin{equation}\label{eq:matrix-rep}
Z_q = \begin{pmatrix}
1 & 0 & 0 & \cdots & 0\\
0 & \omega_q & 0 & \cdots & 0\\
\cdots & \cdots & \cdots & \cdots & \cdots \\
0 & 0 & 0 & \cdots & \omega_q^{q-1}
\end{pmatrix},\qquad X_q = \begin{pmatrix}
  0 & 1 & 0 & \cdots & 0\\
  0 & 0 & 1 & \cdots & 0\\
  \cdots & \cdots & \cdots & \cdots & \cdots \\
  1 & 0 & 0 & \cdots & 0
  \end{pmatrix}.
\end{equation}
These matrices satsify $Z_q^q=X_q^q=\mathbbm{1}$ and the Weyl relation~\cite{weyl1950theory,santhanam1976quantum}
\begin{equation}\label{eq:weyl}
  X_q Z_q = \omega_q Z_q X_q.
\end{equation}
We will drop the $q$ subscript from now on to avoid clutter. If one of the eigenvalues of $Z$ is taken to be 1, the representation \eref{eq:matrix-rep} follows from \eref{eq:weyl}. Denoting the eigenvectors of $Z$ and $X$ as $\ket{z}_Z$ and $\ket{x}_X$ respectively, we have
\begin{equation}
  Z\ket{z}_Z=\omega^z\ket{z}_Z\qquad X\ket{x}_X=\omega^x\ket{x}_X,\qquad z,x\in \mathbb{Z}_q
\end{equation}
The relation between these bases is
\begin{equation}\label{eq:pqrelate}
  \ket{x}_X = \frac{1}{\sqrt{q}}\sum_{z=0}^{q-1} \omega^{zx}\ket{z}_Z,
  \qquad 
  \ket{z}_Z = \frac{1}{\sqrt{q}}\sum_{x=0}^{q-1} \omega^{-zx}\ket{x}_X.
\end{equation}
corresponding to a unitary transformation with the Fourier matrix $F$. The operators satisfy the following relations involving the Fourier matrix :
\begin{equation}\label{eq:fourier-rel}
FZF^\dagger =X ,\qquad FXF^\dagger = Z^{-1}\,.
\end{equation}

The products $Z^a X^b$ with $a,b=0,\ldots q-1$ form a basis of generalized Pauli operators for the $q^2$-dimensional space of operators, and are a quantum mechanical analogue of a classical toroidal phase space, which we expect to recover in the $q\to\infty$ limit. The matrix elements of these products are
\begin{equation}
\left(Z^a X^b\right)_{jk} = \omega^{aj}\delta_{j+b=k\mod q}.
\end{equation}
These transform under the Fourier matrix as
\begin{equation}
F Z^a X^b F^\dagger = X^{a}Z^{-b}\,.
\end{equation}
This shows that unitary transformation by a Fourier transformation is equivalent to a $-\pi/2$ rotation of the toroidal phase space: $(z,x)\longrightarrow (x,-z)$. This will be useful for discussing the classical $q\to\infty$ limit.

\subsection{Cat maps and Clifford gates}

In this section we build up a picture of the Hadamard matrices that are naturally suggested by the Weyl variables, culminating in the cat maps \eref{eq:cat-map}. 

The simplest Hadamard is a power of $Z$ (which is equivalent to a power of $X$ by conjugating with the Fourier matrix c.f. \eref{eq:fourier-rel}). Its effect on a general Pauli operator is
\begin{equation}
Z^k Z^a X^b Z^{\dagger k} = Z^a X^b \omega^{-bk}.
\end{equation}
This corresponds to a `kick' to the momentum $x\to x-k$: just a c-number. As a second example, define $S$ as the diagonal matrix with:
\begin{equation}
S_{jj} = \exp(i\pi j^2/q)\qquad j\in\mathbb{Z}_q.
\end{equation}
This gives
\begin{equation}
S^\alpha Z^a X^b (S^\dagger)^\alpha = \omega^{\alpha/2} Z^{a+\alpha b}X^b\,.
\end{equation}
In this case the momentum has received a kick $\alpha q$, where $q$ is the position. In the phase plane this corresponds to a shear
\begin{align}\label{eq:shear}
  \begin{pmatrix}
    z \\
    x
  \end{pmatrix}&\longrightarrow
  \begin{pmatrix}
    z' \\
    x'
  \end{pmatrix} = \begin{pmatrix}
    z \\
    x + \alpha z
  \end{pmatrix}=
  \begin{pmatrix}
  1 & 0 \\
  \alpha & 1
  \end{pmatrix}
  \begin{pmatrix}
    z \\
    x
  \end{pmatrix}\qquad \mod q.
  \end{align}
The form of $S$ has a simple semiclassical interpretation: after multiplying a wavefunction by $e^{i\theta(z)}$, we identify $\theta'(z)$ with the shift to the momentum. Thus a  shift linear in $z$ corresponds to $\theta(z)\propto z^2$. 

We can combine $S$ with the Fourier map to give the family of Hadamard matrices
\begin{equation}\label{eq:cat-map}
C_{jk}(\alpha,\delta)\equiv\left(S^\alpha F S^\delta\right)_{jk} = \exp\left(\frac{2\pi i}{q}\left[\frac{\alpha j^2}{2} + jk + \frac{\delta k^2}{2}\right]\right)\qquad \alpha,\delta\in \mathbb{Z}.
\end{equation}
These have the following effect on a generalized Pauli matrix 
\begin{equation}\label{eq:1-cat}
Z^a X^b \longrightarrow C(\alpha,\delta)Z^a X^b C^\dagger(\alpha,\delta)=Z^{a'}X^{b'},
\end{equation}
where 
\begin{equation}
\begin{pmatrix}
a' \\
b'
\end{pmatrix}
=\overbrace{\begin{pmatrix}
\alpha & \alpha\delta - 1\\
1 & \delta
\end{pmatrix}}^{\equiv T^T}
\begin{pmatrix}
  a \\
  b
  \end{pmatrix}.
\end{equation}
The unitaries \eref{eq:cat-map} are thus in correspondence with a subfamily of linear area preserving maps of the torus. The general form of such maps is
\begin{align}\label{eq:arnold}
\begin{pmatrix}
  z \\
  x
\end{pmatrix}&\longrightarrow
\begin{pmatrix}
  z' \\
  x'
\end{pmatrix} = T
\begin{pmatrix}
  z \\
  x
\end{pmatrix}\qquad \mod 1\nonumber\\
T &= \begin{pmatrix}
\alpha & \beta \\
\gamma & \delta
\end{pmatrix}\qquad \alpha,\beta,\gamma,\delta\in\mathbb{Z},\qquad \alpha\delta-\beta\gamma=1.
\end{align}
\Eref{eq:cat-map} presents special cases of \emph{quantum cat maps}, first studied in \cite{hannay1980quantization} as the quantum analogs of the classical Arnold cat maps \eref{eq:arnold} \cite{arnol1968problemes}. \Eref{eq:cat-map} corresponds to $\beta=1$, which means they are additionally \emph{Clifford} unitaries i.e. they map Pauli operators to Pauli operators: $Z^a X^b\longrightarrow Z^{a'} X^{b'}$.

The nature of the linear transformation $T$ is determined by $\operatorname{tr} T=\alpha+\delta$: hyperbolic for $|\operatorname{tr} T>2|$, parabolic for $|\operatorname{tr} T|=2$ and elliptic for $|\operatorname{tr} T|<2$. Explicit examples of Clifford quantum cat maps corresponding to the first two are
\begin{align}
  T &= \begin{pmatrix}
    2 & 1 \\
    3 & 2
    \end{pmatrix}: C_{jk}(2,2) = \exp\left(\frac{2\pi i}{q}\left[j^2+k^2+jk\right]\right)\text{ (hyperbolic)}\nonumber\\  
T &= \begin{pmatrix}
-1 & 1 \\
0 & -1
\end{pmatrix}: C_{jk}(-1,-1) = \exp\left(-\frac{i\pi}{q}\left[j-k\right]^2\right)
\text{ (parabolic)}
\end{align}
Note that the second of these leads to a model with $\mathbbm{Z}_q$ \emph{clock symmetry}: $j\longrightarrow j+1 \mod q$ (for $q$ even), which should become a $U(1)$ symmetry in the $q\to\infty$ limit. The parity of $q$ is important since
\begin{equation}
\exp\left(\frac{-i\pi (j+q)^2}{q}\right)=(-1)^q \exp\left(-\frac{i\pi j^2}{q}\right).
\end{equation}
We have not explored the significance of this symmetry any further.

In \cite{moudgalya2019operator}, operator dynamics was studied in the \emph{perturbed} quantum cat maps 
\begin{align}\label{eq:perturbed}
\left(U_q\right)_{jk}= \frac{1}{\sqrt{q}} \exp \left[\frac{2 \pi i}{q \beta}\left(\frac{\alpha j^{2}}{2}+j k+\frac{\delta k^2}{2}\right)+\frac{i \kappa q}{2 \pi} \sin \left(\frac{2 \pi j}{q}\right)\right].
\end{align}
This is still Hadamard for $\beta=1$, but $\kappa\neq 0$ leads to a non-Clifford unitary that gives rise to operator spreading for this single-qudit map. Note that general cat maps are not Hadamard, as some matrix elements vanish \cite{hannay1980quantization}.

According to our general prescription, coupling between sites is introduced via a diagonal matrix with the cat matrix elements down the diagonal
\begin{equation}
C = \operatorname{diag}(C_{j_1j_2}(\alpha,\delta)) \qquad j_{1,2}\in\mathbb{Z}_q\,.
\end{equation}
Ignoring phases\footnote{As we will from now on.} this gives the transformation
\begin{equation}
CZ_1^{a_1} X_1^{b_1} Z_2^{a_2} X_2^{b_2} C^\dagger \sim Z_1^{a_1-\alpha b_1-b_2}X_1^{b_1}Z_2^{a_2-\delta b_2-b_1}X_2^{b_2}\,.
\end{equation}
The momenta on the two sites receive a kick:
\begin{align}\label{eq:2-cat}
z_1&\longrightarrow z_1 \nonumber\\
x_1&\longrightarrow x_1 - z_2 - \alpha z_1\nonumber\\
z_2&\longrightarrow z_2 \nonumber\\
x_2&\longrightarrow x_2 - z_1 - \delta z_2
\end{align}
Combining the single qudit \eref{eq:1-cat} and two qudit \eref{eq:2-cat} maps gives the dynamics of operators in these Clifford circuits, which we turn to now.

\section{Operator dynamics in coupled Clifford cat models} \label{sec:clifford}

Operator spreading in dual-unitary dynamics has been of special interest, where the underlying space-time duality enforces that all operators grow with maximal velocity $|v|=1$~\cite{claeys_maximum_2020,bertini_scrambling_2020,mi_information_2021,rampp_dual_2023,akhtar_dual-unitary_2024}. This will again be the case for all operator dynamics presented below.

The dynamics of these Clifford circuits is specified by finding the evolution of a general Pauli operator, which we write as
\begin{equation}
\mathcal{O}_{AB}\equiv\prod_{x=1}^N Z_x^{a_{x}}X_x^{b_{x}}\qquad a_{x}, b_{x}\in \mathbb{Z}_q,
\end{equation}
where we denote $A=(a_{0}, \ldots, a_{N})$ and $B=(b_{0}, \ldots, b_{N})$. The Clifford property guarantees that the operator remains Pauli at all subsequent times. On conjugation by the unitary describing our model we have $U_t \mathcal{O}_{A_0B_0}U^\dagger_t \equiv \mathcal{O}_{A_t B_t}$, which defines the time evolution of the quantities $A_t=(a_{0,t}, \ldots, a_{N,t})$ and $B_t=(b_{0,t}, \ldots, b_{N,t})$.

Let us take the horizontal bonds to be given by the Fourier matrix $u_\text{H}=F=C(0,0)$ and the vertical bonds to be $u_\text{V}=C(\alpha,\delta)$. The case of cat maps in both directions is not more general because the cat map is the Fourier matrix with a diagonal transformation on either side (see \eref{eq:cat-map}), and so these additional phases can always be placed on the vertical bonds. The result of a horizontal update followed by a vertical update is
\begin{equation}\label{eq:elliptic-cat-hamiltonian}
\begin{aligned}
a_{x,t+1} &= \alpha(a_{x,t}-b_{x-1,t}-b_{x+1,t}) + (\alpha\delta -1)b_{x,t}\\
b_{x,t+1} &= a_{x,t} - b_{x-1,t} - b_{x+1,t}+\delta b_{x,t}
\end{aligned}\mod q.
\end{equation}
These can be regarded as the `Hamiltonian' form of the equations of motion, being first order difference equations involving position and momentum variables. We can eliminate the $a_{x,t}$ to obtain the second difference `Lagrangian' formulation, written in terms of the lattice Laplacian
\begin{equation}
\begin{aligned}
\left[\Delta b\right]_{x,t} &= (\alpha + \delta - 4)b_{x,t}\\
\left[\Delta b\right]_{x,t} &\equiv  b_{x,t+1} + b_{x+1,t-1} + b_{x+1,t} + b_{x-1,t} - 4b_{x,t}
\end{aligned}\mod q
\end{equation}
Here the symmetry between space and time is manifest. If instead the horizontal bonds are taken to be $u_\text{H}=F^\dagger$, the Lagrangian equation involves the lattice d'Alembertian
\begin{equation}\label{eq:F-dagger-eom}
  \begin{aligned}
  a_{x,t+1} &= \alpha(a_{x,t}+b_{x-1,t}+b_{x+1,t}) + (\alpha\delta -1)b_{x,t}\\
  b_{x,t+1} &= a_{x,t} + b_{x-1,t} + b_{x+1,t}+\delta b_{x,t}\\
  \left[\square b\right]_{x,t} &= (\alpha + \delta)b_{x,t}\\
  \left[\square b\right]_{x,t} &\equiv b_{x,t+1} + b_{x+1,t-1} - b_{x+1,t} - b_{x-1,t}
  \end{aligned}\mod q
\end{equation}
The operator dynamics of the two models may be simply related by changing the sign of $a_{x,t}$, $b_{x,t}$ on alternate sites as $a_{x,t}\rightarrow (-1)^x a_{x,t}$, and $b_{x,t}\rightarrow (-1)^x b_{x,t}$.

Note that since the equations of motion are linear, it suffices to understand the dynamics (assuming translational invariance) starting from a single initial $Z$ as well as from a single $X$. The general case can then be obtained by adding these solutions to satisfy any desired initial condition.

The models introduced in this section have appeared in several different contexts --- with several different names --- over the years. They are an example of \emph{Clifford cellular automata} (CCAs) \cite{schlingemann2008structure}, a particular class of quantum cellular automata (see \cite{farrelly2020review} for a recent review). Strictly CCA dynamics keeps track of the phase of the operators: by discarding the phase (which can always be reconstructed) we have arrived at linear dynamics on the discrete torus which preserves the symplectic inner product, sometimes described as a \emph{symplectic cellular automaton} (see \cite{kaneko1988symplectic} for an earlier related construction). Finally, the classical version of this model was described as a \emph{coupled cat map model} \cite{gutkin2016classical} or \emph{spatiotemporal cat} \cite{gutkin2021linear}. These works also included an on-site potential akin to the perturbed cat maps \eref{eq:perturbed} that can give rise to non-Clifford dynamics and operator entanglement. We note that all ideas generalize readily to higher dimensional situations, e.g. for a model defined on a cubic lattice.

\subsection{Fractal dynamics: $\alpha+\delta\neq 0 \mod q$}

Ref.~\cite{gutschow2010time} describes an automaton in terms of a characteristic dynamical matrix, essentially using the additive structure and translational invariance to obtain the equations of motion for the variables $a_t(u)\equiv\sum_x a_{x,t}u^x$ and $b_t(u)\equiv\sum_x b_{x,t}u^x$, with
\begin{equation}
\begin{pmatrix}
a_{t+1}(u) \\
b_{t+1}(u)
\end{pmatrix}
= M(u,u^{-1})
\begin{pmatrix}
  a_{t}(u) \\
  b_{t}(u)
  \end{pmatrix}
\end{equation}
For the equations of motion \eref{eq:F-dagger-eom}, choosing $u_\text{H}=F^\dagger$ in order to make the connection with previous results, we have
\begin{equation}
 M(u, u^{-1}) = \begin{pmatrix}
 \alpha & (\alpha\delta - 1) + \alpha(u + u^{-1}) \\
 1 & \delta + u + u^{-1}
 \end{pmatrix}
\end{equation}
The trace is $\operatorname{tr}M = \alpha + \delta + u + u^{-1}$. According to the results of \cite{gutschow2010time} this means that for $\alpha+\delta=0$ we have a glider automaton (i.e. one with ballistically propagating excitations, see next subsection) and a fractal automaton otherwise. This dependence on $\alpha+\delta$ is already apparent in the equations of motion \eref{eq:F-dagger-eom}, showing that the second order dynamics of the $b_{x,t}$ depends only on the sum $\alpha+\delta$ (also implying that $\alpha+\delta=0$ is not fundamentally different from $\alpha=\delta=0$). 

\begin{figure}
  \begin{center}
  \includegraphics[width=0.48\textwidth]{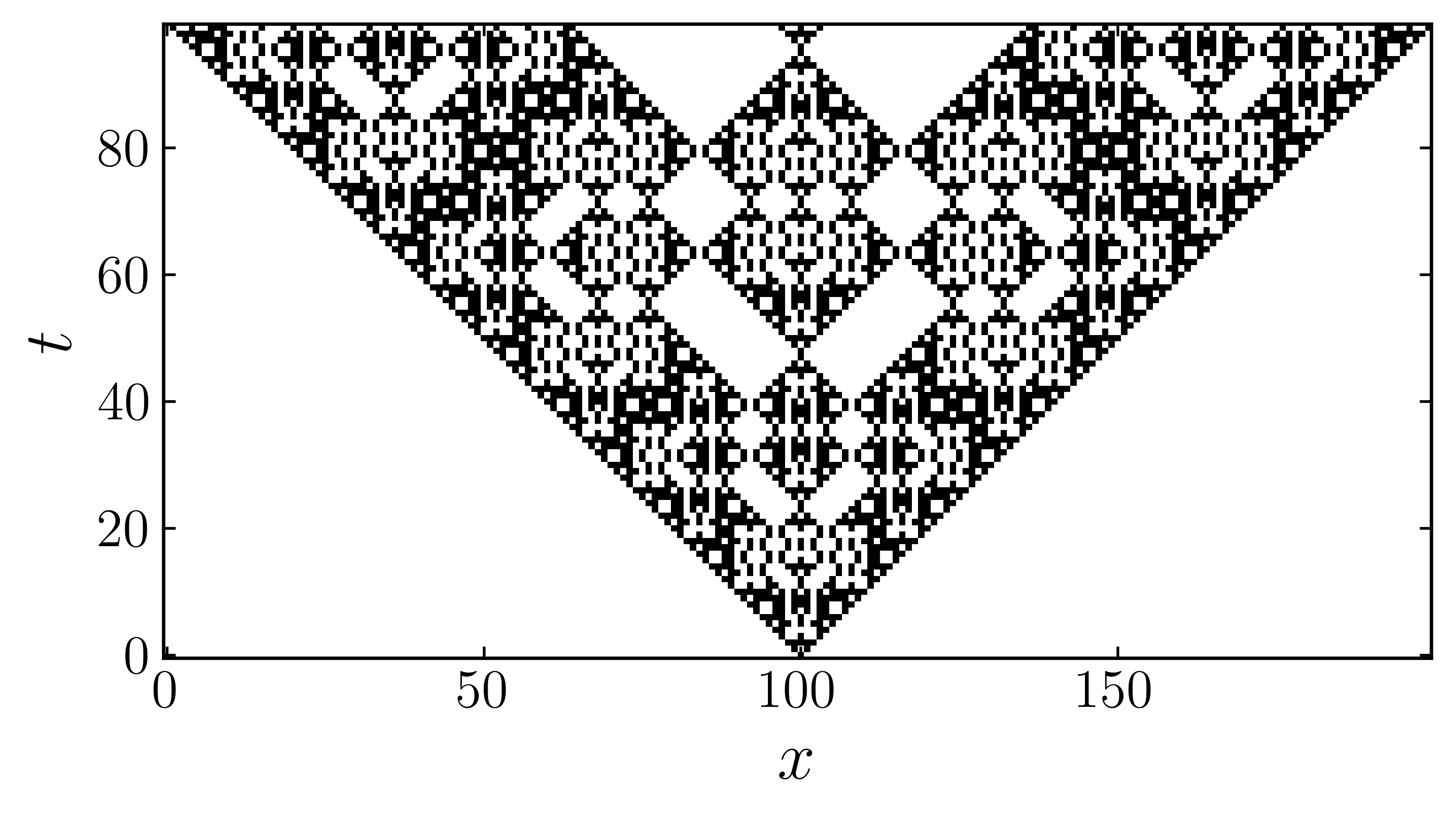}
  \includegraphics[width=0.48\textwidth]{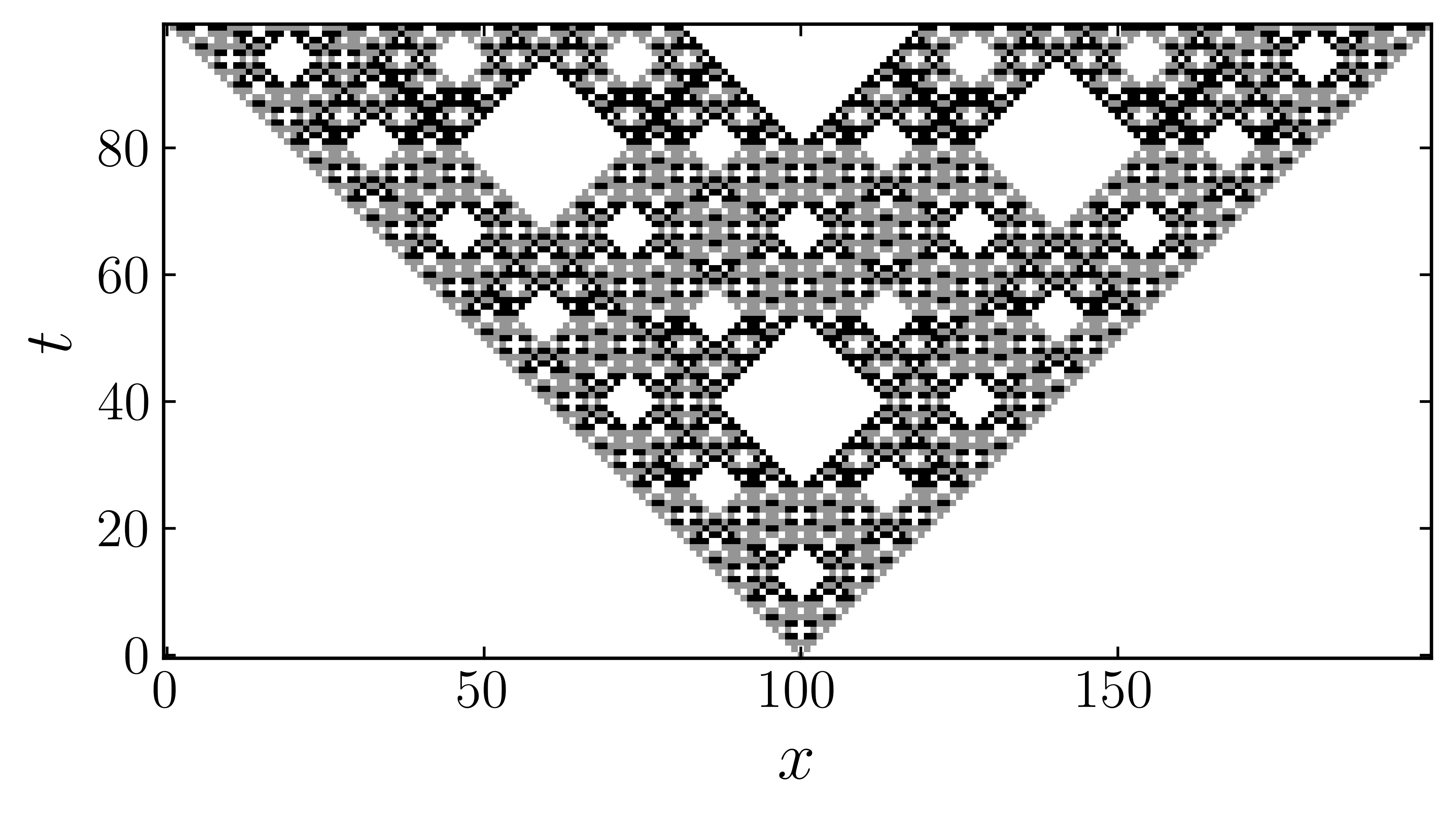} 
  \caption{Operator dynamics for $v_\text{H}=F^\dagger$, $u_\text{V}=C(1,0)$. The figures illustrate the coefficients $b_{x,t}$ for propagation starting from a single $X$ for $d=2$ (left), the kicked Ising model in a longitudinal field $h=\pi/4$ \cite{gutschow2010time,sommers2023crystalline} and the $d=3$ model of \cite{lotkov2022floquet} (right). On the right white, grey, and black represent $X^0=\mathbbm{1}$, $X$, and $X^2$.}
  \label{fig:fractal}
  \end{center}
\end{figure}

In the generic case of $\alpha+\delta \neq 0$ the operator dynamics hence generally exhibits a fractal structure. This fractal structure of abelian automata –- of which the Clifford cellular automata are an example -- is described in \cite{gutschow2010fractal}. Such fractal structures already appeared in the literature: The couplings $u_\text{H}=F^\dagger$, $u_\text{V}=C(1,0)$ give the patterns discussed in \cite{gutschow2010time,gutschow2010fractal,sommers2023crystalline} for $q=2$, and in \cite{lotkov2022floquet} for $q=3$, see \fref{fig:fractal}. Similar Clifford dynamics underlie the fractal patterns in
\cite{gopalakrishnan2018facilitated,klobas_exact_2023,sommers_zero-temperature_2024}.

\subsection{Glider dynamics: $\alpha=\delta = 0\mod q$} \label{sec:no-cat}

For $\alpha=\delta=0$ we expect to recover a glider automaton. This can be made apparent starting from the equations of motion for $u_\text{H}=F^\dagger$, which take the simple form
\begin{equation}\label{eq:glider-eom}
\begin{aligned}
  a_{x,t+1} &=  -b_{x,t}\nonumber\\
  b_{x,t+1} &= a_{x,t} + b_{x-1,t} + b_{x+1,t}
\end{aligned} \mod q.
\end{equation}
Gliders can be directly obtained by noting that these equations support solutions as configurations translating to the left or right,
\begin{align}
a^R_{x,t} &= r_{x-t}, \qquad b^R_{x,t} = -r_{x-t-1}\,,\nonumber\\
a^L_{x,t} &= l_{x+t}, \qquad b^L_{x,t} = -l_{x+t+1}\,,
\end{align}
which is also evident from the second difference form $\left[\square b\right]_{x,t} = 0$. Since the equations of motion are linear, we can add these solutions mod $q$ to obtain new solutions. 

The simplest propagating solutions are
\begin{align}\label{eq:two-site}
&\text{right moving: } Z_j X^{-1}_{j+1}\nonumber\\
&\text{left moving: } X^{-1}_{j}Z_{j+1} .
\end{align}
Adding the solutions corresponds to taking products of two-site operators, such that e.g. $Z_j^n X^{-n}_{j+1}$ and $X^{-n}_{j}Z^n_{j+1}$ are also right and left propagating solutions respectively. In this way we obtain $2$ sets of $q$ independent gliders per lattice set $j$, returning a complete (operator) basis and indicating that any initial operator can be decomposed in left- and right-moving gliders. For a system of size $N$ this implies a recurrence time of $N$, since after $N$ time steps and corresponding translations any initial operator will return to itself. Such linear recurrence times were already obtained in \cite{gombor_superintegrable_2022,peng_quantum_2022} for dual-unitary models where the underlying gates additionally satisfy the set-theoretic Yang-Baxter equation and square to the identity, a point we will return to in \sref{sec:yb}.

Note that a single-site Pauli operator leads to a more complicated evolution since all gliders are two-site operators. For example, starting from a single $X$ operator at $x,t=0$ gives a checkerboard wedge
\begin{align}
  a_{x,t} &= \begin{cases}
    -1 & |x|\leq t,\qquad  x + t = 1 \mod 2\\
    0 & \text{ otherwise}
\end{cases}
\nonumber\\ 
b_{x,t} &= \begin{cases}
1 & |x|\leq t,\qquad  x + t = 0 \mod 2\\
0 & \text{ otherwise}
\end{cases}
\end{align}
All single-site correlation functions therefore vanish. The two-site correlation function corresponding to the solutions \eref{eq:two-site} are nonzero and constant along a light cone.

Finally, note that the equations of motion \eref{eq:glider-eom} leave invariant the `vacuum' states $a_{x,t}=-b_{x,t}$ This gives rise to the following propagating solutions
\begin{align}\label{eq:parafermion_prop}
\Psi_X(j)&\equiv\left(\prod_{k<j} Z_k X^{-1}_k\right) X^{-1}_j\qquad \text{to the right}\nonumber\\
\Psi_Z(j)&\equiv\left(\prod_{k<j} Z_k X^{-1}_k\right) Z_j\qquad \text{to the left}.
\end{align}
Note that the above elementary glider solutions correspond to 
\begin{align}
Z_jX^{-1}_{j+1}=\Psi^{-1}_X(j)\Psi_X(j+1)\nonumber\\
Z_j^{-1}Z_{j+1}=\Psi^{-1}_Z(j)\Psi_Z(j+1),
\end{align}
moving to the right and left respectively. 

These operators obey the relations $\Psi_X^q(j) = \Psi_Z^q(j) = 1$ as well as
\begin{align}
  \Psi_X(j)\Psi_{X}(k) &= \omega^{\sign(j-k)}\Psi_{X}(k)\Psi_X(j)\nonumber\\
  \Psi_Z(j)\Psi_{Z}(k) &= \omega^{\sign(k-j)}\Psi_{Z}(k)\Psi_Z(j)\nonumber\\
  \Psi_X(j)\Psi_{Z}(k) &= \omega^{\sign(j-k)}\Psi_{Z}(k)\Psi_X(j)\qquad j\neq k\nonumber\\
\Psi_Z(j)\Psi_{X}(j) &= \omega\Psi_{X}(j)\Psi_Z(j).
\end{align}
These are therefore \emph{parafermion} operators that generalize the Jordan--Wigner fermions of the Ising model to $q>2$ \cite{fradkin1980disorder,fendley2012parafermionic}. The ballistically propagating solutions \eref{eq:parafermion_prop} are the analog for unitary dynamics of the discrete holomorphic conditions in statistical mechanical models discussed in \cite{rajabpour2007discretely}.

\subsection{Dynamics of product states}

We now revisit the cellular automata picture discussed in \sref{sec:clockwork}, which corresponds exactly to the $\alpha=\delta = 0$ case discussed in the previous subsection. Recall that by summing over one sublattice we found the second order dynamics \eref{eq:second-order} on the states of the other sublattice. 

More precisely, we consider starting from a product state of alternating $Z$ and $X$ eigenstates
\begin{equation}\label{eq:initial-product}
\ket{z_1}_Z\ket{x_2}_X\cdots \ket{z_{N-1}}_Z\ket{x_{N}}_X \,,\quad\qquad z_{2j-1}, x_{2j} \in\mathbb{Z}_q \, \text{ for } \, j=1,\ldots N/2.
\end{equation}
Taking $u_\text{V}=u_\text{H}=F$ and applying $U_\text{vert}U_\text{row}$ to this state gives
\begin{equation}\label{eq:final-product}
  \ket{z_1}_X\ket{-x_2-z_1-z_3}_Z\cdots \ket{z_{N-1}}_X\ket{-x_N-z_{N-1}-z_1}_Z
\end{equation}
(as usual addition is modulo $q$), and similarly for the state $\ket{x_1}_X\ket{z_2}_Z\cdots \ket{x_{N-1}}_X\ket{z_{N}}_Z$. In this basis the dynamics explicitly corresponds to a cellular automaton. We can view this graphically as follows. The state \eref{eq:initial-product} has the representation
\begin{equation}
  \vcenter{\hbox{\includegraphics[width=0.36\linewidth]{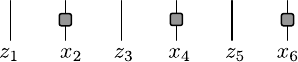}}}\,,
\end{equation}
since the $X$ basis is related to the $Z$ basis through a unitary transformation with the Fourier matrix. 
After applying $U_\text{vert}U_\text{row}$ we have
\begin{align}
  \vcenter{\hbox{\includegraphics[width=0.4\linewidth]{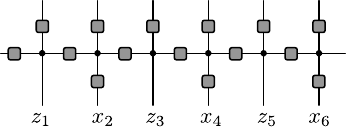}}}
  \quad= \quad\vcenter{\hbox{\includegraphics[width=0.4\linewidth]{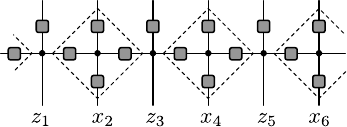}}}.
\end{align}
In this equality we have made explicit the dual-unitary interactions round-a-face of \eref{eq:clockworkgate}. Since the diamond regions correspond to the application of Rule 150R (as discussed in \sref{sec:clockwork}), the remaining Hadamards on the odd sites give the $XZXZ\cdots$ pattern of eigenstates in \eref{eq:final-product}. Applying $U_\text{vert}U_\text{row}$ a second time
\begin{align}
  \vcenter{\hbox{\includegraphics[width=0.4\linewidth]{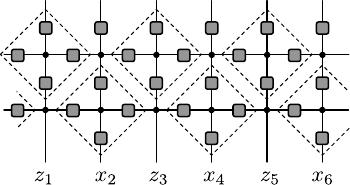}}}
  \end{align}
reproduces the original $ZXZX\cdots$ pattern.

The relation to the Clifford dynamics described by \eref{eq:elliptic-cat-hamiltonian} for $\alpha=\delta=0$ is as follows. In one time step we have the operator dynamics
\begin{align}
Z_{2j-1}\longrightarrow X_{2j-1}, \qquad
X_{2j}\longrightarrow X_{2j-1}^{-1} Z_{2j}^{-1} X_{2j+1}^{-1}.
\end{align}
The initial state \eref{eq:initial-product} is an eigenvector of $Z_{2j-1}$ and $X_{2j}$ for $j=1,\ldots N/2$ with eigenvalues $\omega^{q_{2j-1}}$ and $\omega^{p_{2j}}$ respectively. It is straighforward to check that the final state \eref{eq:final-product} is an eigenstate of $X_{2j-1}$ and $X_{2j-1}^{-1} Z_{2j}^{-1} X_{2j+1}^{-1}$ with the same eigenvalues. The same applies to any products of the above operators, since they are mutually commuting, and hence to multiple time steps.

Mapping product states of the form \eref{eq:initial-product} to product states means that this initial condition generates no entanglement. On the other hand, it is known that for the self-dual kicked Ising model, which is in the class we study here, an initial product state consisting of only $X$ or only $Z$ eigenstates generates maximal entanglement \cite{bertini2019entanglement}. To be precise, consider the bipartition of an infinite system into two semi-infinite halves $A$ and $\bar A$, with the initial state a product of $Z$ eigenstates. After $T\geq 1$ time steps, the reduced density matrix  of one of these halves has $q^{T-1}$ eigenvalues all equal to $q^{-T+1}$, with the remainder zero. Then the Rényi entanglement entropies obey
\begin{equation}
S_A^{(\alpha)}(T) \equiv \frac{1}{1-\alpha} \log\tr\left[\rho_A^\alpha\right] = (T-1) \log q,
\end{equation}
which is maximal\footnote{For a product state of $X$ eigenstates the result is $T \log q$.}. To understand the origin of this behaviour in the present context, we express a product state in the $Z$ eigenbasis in terms of the states \eref{eq:initial-product} (c.f. \eref{eq:pqrelate})
\begin{align}\label{eq:qprod-expansion}
  &\ket{z_1}_Z\ket{z_2}_Z\cdots \ket{z_{N-1}}_Z\ket{z_{N}}_Z \nonumber\\
  &\qquad= \frac{1}{q^{N/2}}\sum_{x_2,x_4,\ldots x_N}\omega^{-z_2x_2-z_4x_4-\cdots z_N x_N}\ket{z_1}_Z\ket{x_2}_X\cdots \ket{z_{N-1}}_Z\ket{x_{N}}_X.
\end{align}
The unitary evolution of this state for $T$ timesteps can be understood in terms of the Rule 150R dynamics acting on the states \eref{eq:initial-product}. This causes each index $x_2,x_4 \ldots x_N$ that appears in \eref{eq:qprod-expansion} to appear in multiple factors of the evolved product state, with the index $x_{2j}$ appearing in sites in the forward light cone of the site $2j$, namely the sites $[2j-T+1,2j+T-1]$ (for $T\geq 1$)
\begin{equation}
  \vcenter{\hbox{\includegraphics[width=0.4\linewidth]{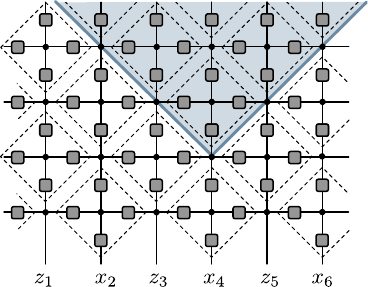}}}.
\end{equation}
The sum over $x_{2j}$ in \eref{eq:qprod-expansion} then contributes to the entanglement between two regions that intersect with this light cone. For the case of two semi-infinite halves mentioned above, $T-1$ indices appear in both halves. Since the Rule 150R dynamics is bijective, every term in \eref{eq:qprod-expansion} gives rise to a distinct state. Since all coefficients in the resulting expansion have the same magnitude, the entanglement entropies are maximal and equal to $(T-1)\log q$.

A slightly more general result is obtained if, in the expansion \eref{eq:qprod-expansion}, we weight each state $\ket{x}_X$ with an amplitude $c_x$ satisfying $\sum_{x=0}^{q-1} |c_x|^2=1$ instead of $q^{-1/2}$. In this case the above argument still goes through, giving the von Neumann entanglement entropy (corresponding to Rényi index 1)
\begin{equation}
  S_A^{(\alpha)}(T)= -T \sum_{x=0}^{q-1} |c_x|^2\log |c_x|^2.
\end{equation}
This allows the rate of entanglement growth (the entanglement velocity) to be tuned from zero to the maximal value $(T-1)\log q$. This is reminiscent of the numerical results of \cite{bertini2019entanglement} for the self-dual kicked Ising model in the integrable (zero longitudinal field) limit and the analytical results of \cite{klobas2021exact} on the Rule 54 automaton.

This result, holding for $\alpha+\delta=0$ and a very special set of initial conditions, does not invalidate the general results connecting dual unitarity to maximal entanglement growth in \cite{zhou_maximal_2022,piroli_exact_2020,foligno_growth_2023}. In particular, the last paper shows that initial states written as products of arbitrary two-site states (which includes the states considered here) have maximal entanglement growth for \emph{almost all} dual unitary circuits. The model considered here is presumably a member of a set of exceptions of measure zero.

\section{Integrable Potts models, solitons, and long-range entanglement generation} \label{sec:lotkov}

In the previous section it was shown that the operator dynamics can support propagating solutions, also known as gliders, reminiscent of the solitons underlying integrability. 
In a recent work~\cite{lotkov2022floquet}, Lotkov \emph{et al.} considered a particular Floquet dynamics in a $q=3$ chain that turns out to be an example of the models considered here. The authors considered a kicked Potts model, and at specific points in parameter space it was shown that the model supports specific protocols to generate long-range entanglement in the form of ``rainbow states'' starting from an initial product state. It was conjectured that this model is integrable, which was subsequently established in \cite{miao2023integrable}. Other points in the parameter space correspond to the fractal Clifford dynamics illustrated in \fref{fig:fractal}.

Here we show how this model can be naturally rewritten as a Hadamard circuit, with the special property that the Hadamard matrices on the horizontal links are the inverse of those on the vertical links. For such circuits we identify gliders from which an extensive set of local conserved charges can be constructed in an alternative and purely graphical way.
We similarly present a purely graphical derivation for the protocol generating long-range entanglement. Interpreting the dynamics as unitary circuit dynamics, we show how the underlying gates satisfy the set-theoretical Yang-Baxter relation, although curiously only for $q<6$ or if the dynamics correspond to Clifford dynamics. This approach further highlights the ways in which this model extends the self-dual kicked Ising model for the $q=2$ chain, and directly extends these constructions to arbitrary $q$ and away from the Clifford points. 

\subsection{From kicked Potts models to lattice models}

Fixing $d=3$, the Floquet protocol of \cite{lotkov2022floquet} considered kicked Floquet dynamics generated by two Hamiltonians\footnote{In order to make the connection with Hadamard circuits more transparent, we here exchange the use of $X$ and $Z$ as compared to the original reference, which corresponds to a local basis transformation generated by the Fourier matrix.},
\begin{align}
H_1 = \sum_{j=1}^{2N-1} Z_j^2 Z_{j+1} + Z_{j}Z_{j+1}^2, \qquad H_{2} = \sum_{j=1}^{2N} \left(X_j+X_j^2\right),
\end{align}
where $Z_j$ ($X_j$) now denotes the operator $Z$ ($X$) acting on site $j$. Evolving the system with Hamiltonian $H_1$ for a time $T_1$ and Hamiltonian $H_2$ for a time $T_2$, we can identify the corresponding unitary dynamics with a tensor network of the form in Fig.~\ref{fig:tn}, with
\begin{align}
U_{\text{row}} = e^{-i H_1 T_1}, \qquad  U_{\text{vert}} = e^{-i H_2 T_2}\,.
\end{align}
Here we used that the first Hamiltonian is diagonal in the computational basis, whereas the second Hamiltonian consists of one-site unitaries. This model was shown to be integrable for $T_1 = T_2 = \alpha$, where $\alpha = \frac{2\pi}{9}(3l-m)$ with $m =1,2$ and $l\in \mathbb{Z}$ ($m=0$ results in trivial dynamics with the unitaries reducing to the identity). For general $\alpha$ it follows that
\begin{align}
u_{\text{H}}(z_i,z_j) = \delta_{z_i,z_j} e^{-2i \alpha}+ (1-\delta_{z_i,z_j})e^{i\alpha}
\end{align}
and
\begin{align}
u_{\text{V}}(z_i,z_j) = \frac{1}{3}\left(e^{-2i\alpha}-e^{i\alpha}\right)+\delta_{z_i,z_j} e^{i\alpha}\,.
\end{align}
At the integrable point with $m=1$ we find that
\begin{align}
u_{\text{H}}(z_i,z_j) = e^{i\frac{4\pi}{9}} K_3^{\dagger}(z_i,z_j), \qquad u_{\text{V}}(z_i,z_j) = -i\frac{e^{i\frac{4\pi}{9}}}{\sqrt{3}}K_3(z_i,z_j),
\end{align}
whereas for $m=2$ we recover
\begin{align}
u_{\text{H}}(z_i,z_j) = e^{i\frac{8\pi}{9}} K_3(z_i,z_j), \qquad u_{\text{V}}(z_i,z_j) = i\frac{e^{i\frac{8\pi}{9}}}{\sqrt{3}}K_3^{\dagger}(z_i,z_j).
\end{align}
Up to an unimportant prefactor, all matrices are (proportional to) complex Hadamard matrices. Furthermore, in both models we recover the setup from Fig.~\ref{fig:tn}, with the additional property that the complex Hadamard matrices on the vertical bonds are the hermitian conjugate of those on the horizontal bonds. We will graphically represent such circuits as
\begin{equation}\label{eq:integrablecircuit}
\vcenter{\hbox{\includegraphics[width=0.45\linewidth]{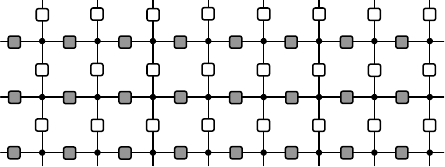}}} \qquad 
\vcenter{\hbox{\includegraphics[width=0.41\linewidth]{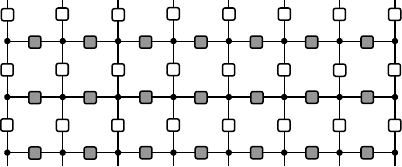}}}\,,
\end{equation}
for periodic and closed boundary conditions respectively. The gray and white squares represent symmetric complex Hadamard matrices that are each others inverse. It is a direct check that the conditions for the dynamics to correspond to a Hadamard circuit is equivalent to the integrability constraints. For local qubits with $d=2$, this setup is also realized in the self-dual kicked Ising model at the integrable point (see e.g.~\cite{claeys2023dual}). For this model all complex Hadamard matrices are identical and equal to $F_2$, which satisfies $F_2^{\dagger} = F_2$, such that the above setup directly applies.

In the following, we will consider models of the form~\eqref{eq:integrablecircuit} for more general complex Hadamard matrices. The only requirements are that (i) the complex Hadamard matrices are symmetric, and (ii) $u_{\textrm{H}}^{\dagger} = u_{\text{V}}$. If we consider complex Hadamard matrices that are equivalent to the Fourier matrix, both the phases and permutations exactly cancel at each vertex due to the geometry of the lattice and the model reduces to Clifford dynamics equivalent to a classical cellular automaton, but the construction does not depend on any specific parametrizations and supports more general dynamics.

\subsection{Long-range entanglement generation}
These circuits can be used to generate long-range entanglement starting from an initial product state. The explicit decomposition in terms of complex Hadamard matrices makes clear that these protocols directly generalize to any circuit composed out of symmetric complex Hadamard matrices for which $u_{\textrm{H}}^{\dagger} = u_{\text{V}}$ and significantly simplifies the corresponding derivation, which can be done purely graphically.  The generalization of the proposed protocol from \cite{lotkov2022floquet} consists of three steps.

\begin{itemize}
\item Prepare the system in an initial product state $\ket{\psi(t=0)} = \otimes_{j=1}^{2N} \ket{+}_j$, where $\ket{+} = \frac{1}{\sqrt{q}}\sum_{a=1}^q \ket{a}$.
\item Evolve the initial state for $N$ discrete time steps with an evolution operator with closed boundary conditions in which the horizontal unitary between sites $N$ and $N+1$ is removed (i.e. there is no interaction between sites $N$ and $N+1$).
\begin{align}
[U_{\text{vert}}\tilde{U}_{\text{row}}]^N\ket{\psi(t=0)}, 
\end{align}
with the row unitary defined as
\begin{align}
\braket{z_{1:2N}|\tilde{U}_{\text{row}}|z_{1:2N,t}} = \prod_{j=1}^{N-1} u_\text{H}(z_{j,t},z_{j+1,t})\prod_{j=N+1}^{2N-1} u_\text{H}(z_{j,t},z_{j+1,t})\,.
\end{align}
\item Evolve the resulting state for $N$ discrete time steps with the inverse of the full unitary evolution, in which the interaction between sites $N$ and $N+1$ is restored.
\begin{align}
\ket{\psi_{\text{final}}} = [U_{\text{vert}}U_{\text{row}}]^{\dagger N}[U_{\text{vert}}\tilde{U}_{\text{row}}]^N\ket{\psi(t=0)}.
\end{align}
The resulting state is given by a rainbow state, in which the sites $N-j+1$ and $N+j$ are connected by a maximally entangled Bell-like state,
\begin{align}
\ket{\psi_{\text{final}}} = \bigotimes_{j=1}^N \ket{\psi_K}_{N-j-1,N+j}, \qquad \ket{\psi_K} = \frac{1}{\sqrt{q}}\sum_{a,b=1}^q u_{\text{H}}(a,b)\ket{a}\otimes\ket{b}\,.
\end{align}
The unitarity of $u_{\text{H}}$ directly implies that the two sites are maximally entangled.

\end{itemize}

In order to graphically represent this protocol, we can build up the full dynamics starting from the initial state, which can be conveniently represented as
\begin{align}
\otimes_{j=1}^{2N} \ket{+}_j =\,\,  \vcenter{\hbox{\includegraphics[width=0.31\linewidth]{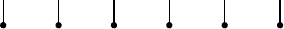}}}\,\,,
\end{align}
taking $N=3$ for concreteness. The full circuit representation of the final state then follows as 
\begin{align}
 \vcenter{\hbox{\includegraphics[width=0.5\linewidth]{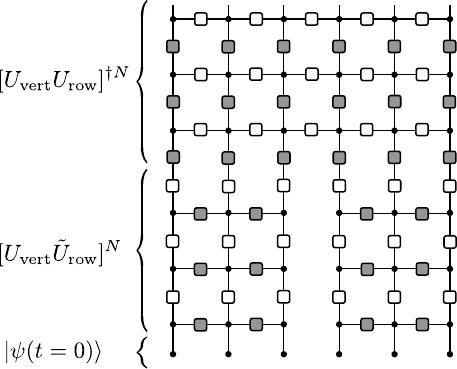}}}\, = \, \vcenter{\hbox{\includegraphics[width=0.31\linewidth]{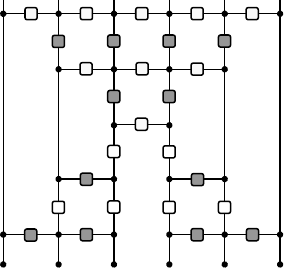}}} 
\end{align}
In order to simplify the circuit, we note that many of the unitaries in $[U_{\text{vert}}U_{\text{row}}]^{\dagger N}[U_{\text{vert}}\tilde{U}_{\text{row}}]^N$ cancel due to unitarity, introducing a `causal light cone' starting from sites $N,N+1$.

Next, we note that the initial state can effectively be `absorbed' in the circuit, since the contraction of two delta tensors returns a delta tensor. This procedure is identical to the so-called `spider fusion' of ZX calculus~\cite{van_de_wetering_zx-calculus_2020}, which similarly concerns compositions of delta tensors, leading to
\begin{align}
 \vcenter{\hbox{\includegraphics[width=0.31\linewidth]{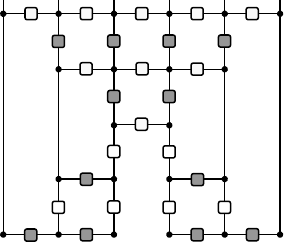}}} \, = \,  \vcenter{\hbox{\includegraphics[width=0.31\linewidth]{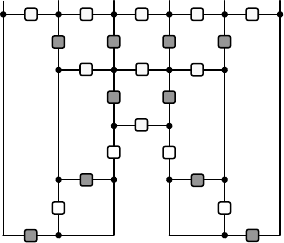}}}
\end{align}
In this equality we have used that the unitary matrices on the vertical legs are the hermitian conjugate of the horizontal ones, and identified the places where we multiply $u_{\textrm{H}}$ with $u_{\textrm{V}}$. Next we can use that the matrix elements are phases, such that any places where the delta tensors introduce a term $u_{\textrm{H}}(a,b)u_{\textrm{V}}(a,b)=|u_\textrm{H}(a,b)|^2=1$ can also be removed:
\begin{align}
 \vcenter{\hbox{\includegraphics[width=0.31\linewidth]{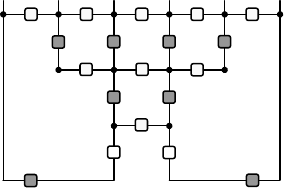}}} \, = \,  \vcenter{\hbox{\includegraphics[width=0.31\linewidth]{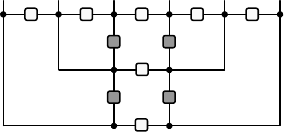}}}
\end{align}
where we have again simplified the circuit by contracting vertical with horizontal unitaries. Repeating these two steps, either cancelling phases or contracting the unitary matrices, we find that
\begin{align}
 \vcenter{\hbox{\includegraphics[width=0.31\linewidth]{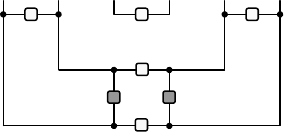}}} \, = \,  \vcenter{\hbox{\includegraphics[width=0.31\linewidth]{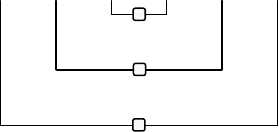}}}
\end{align}
This final expression results in a rainbow state where all sites $N-j+1$ and $N+j$ are connected by a unitary operator $u_{\text{H}}$ resulting in maximally entangled Bell-like states.
\begin{align}
\ket{\psi_{\text{final}}} =\bigotimes_{j=1}^N \left(\frac{1}{\sqrt{q}}\sum_{a,b=1}^q u_{\text{H}}(a,b)\ket{a}_{N-j-1}\ket{b}_{N+j}\right)
\end{align}
The previous derivation implicitly used that $u_{\text{H}}$ and hence $u_{\text{V}}$ were symmetric. This is guaranteed whenever the complex Hadamard matrices are Fourier matrices, and also holds for $K_3$ and $F_4^{(1)}(a), \forall a$, but is not generally the case.

\subsection{Conserved charges and integrability}
Integrable Clifford dynamics are characterized by the presence of gliders, propagating as solitons throughout the systems. Similarly, integrability in dual-unitary circuits implies the presence of solitons propagating with maximal velocity~\cite{bertini_operator_2020,holden-dye_fundamental_2023}. This demand is more stringent than the usual integrability conditions, i.e. the existence of an extensive set of conserved charges, but such solitons can be directly used to construct conserved charges as sums of local operators.

In this section we explicitly construct these solitons and the resulting conserved charges for lattice dynamics of the form~\eqref{eq:integrablecircuit}. It is instructive to first consider the corresponding brickwork circuit, with local gates $U$ of the form
\begin{equation}
\vcenter{\hbox{\includegraphics[width=0.28\linewidth]{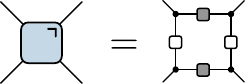}}}\,\,.
\end{equation}
For these gates, the $\ket{+} = \frac{1}{\sqrt{q}}\sum_{a=1}^q \ket{a}$ states behave as solitons. The unitary gates $U$ act as swap gates whenever one of the two states is the $\ket{+}$ state, i.e. $U(\ket{+}\otimes \ket{\psi}) = \ket{\psi} \otimes \ket{+}$ and  $U  (\ket{\psi} \otimes \ket{+}) =  \ket{+} \otimes \ket{\psi}$ for all states $\ket{\psi}$. The derivation works similarly as for the entanglement generation protocol, e.g. for $U \ket{+}\otimes \ket{\psi}$ we can use that:
\begin{equation}\label{eq:Rmatrix_plus}
\vcenter{\hbox{\includegraphics[width=0.1\linewidth]{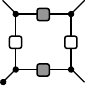}}}
=\vcenter{\hbox{\includegraphics[width=0.1\linewidth]{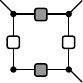}}}
=\vcenter{\hbox{\includegraphics[width=0.1\linewidth]{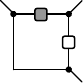}}}
=\vcenter{\hbox{\includegraphics[width=0.1\linewidth]{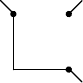}}}
=\vcenter{\hbox{\includegraphics[width=0.1\linewidth]{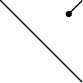}}}
\end{equation}
Conserved charges for the unitary circuit dynamics can be directly constructed from projectors on the $\ket{+}$ state. However, because of the different boundary conditions in the brickwork circuits and the coupled cat models these do not directly translate to solitons for the coupled cat model. Informally, the propagation of solitons needs an initial `corner' to start the contractions, which is absent in the circuits~\eqref{eq:integrablecircuit}.
 
Solitons for the Hadamard models can be introduced as two-site operators [similar to \eqref{eq:two-site}] that cancel an initial horizontal Hadamard matrix and perform an appropriate contraction. Propagating solutions are given by
\begin{align}\label{eq:Had_gliders}
&\text{right moving: } g_{j}^{(+)} = \sum_{a,b,c=1}^q u^*_\text{H}(a,b) u_\text{H}(a,c) \, (\ket{a}\bra{a})_j (\ket{b}\bra{c})_{j+1}\nonumber\\
&\text{left moving: }  g_{j}^{(-)} =  \sum_{a,b,c=1}^q u^*_\text{H}(a,b) u_\text{H}(a,c) \, (\ket{b}\bra{c})_j (\ket{a} \bra{a})_{j+1}
\end{align}
These can again be analyzed purely graphically. Consider the right-moving solutions, graphically represented as
\begin{equation}
\vcenter{\hbox{\includegraphics[width=0.42\linewidth]{fig_rglider_0}}}
\end{equation}
Under a unitary transformation with $U_{\text{vert}} U_{\text{row}}$, this operator transforms as
\begin{equation}
\vcenter{\hbox{\includegraphics[width=0.5\linewidth]{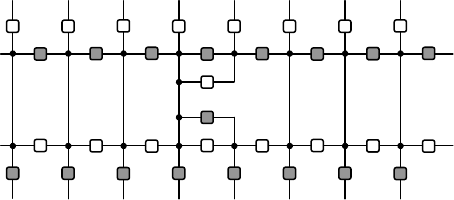}}} \quad  = \quad \vcenter{\hbox{\includegraphics[width=0.42\linewidth]{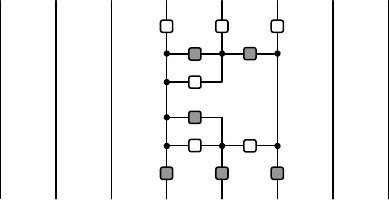}}} \,,
\end{equation}
where we have again used unitarity to contract all operators outside of the causal light cone of the operator. Using the properties of the complex Hadamard matrices, we can first cancel opposite phases before using the unitarity of the matrices to obtain
\begin{align}
 \vcenter{\hbox{\includegraphics[width=0.42\linewidth]{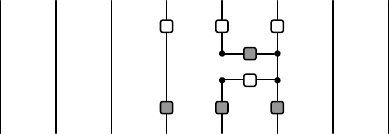}}} \quad = \quad \vcenter{\hbox{\includegraphics[width=0.42\linewidth]{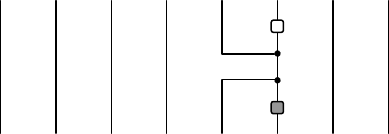}}} \,.
\end{align}
This final expression can be directly deformed to return the original operator shifted by a single site to the right
\begin{equation}
\vcenter{\hbox{\includegraphics[width=0.42\linewidth]{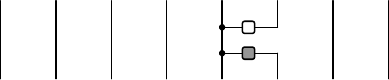}}}
\end{equation}
The connection with the states $\ket{+}$ acting as solitons for the unitary gates can be directly made by observing that these gliders are a unitary transformation of $\mathbbm{1} \otimes \ket{+}\bra{+}$ with a two-site horizontal bond gate $u_{\textrm{H}}$:
\begin{align}
u_{\textrm{H}}^{\dagger}(\mathbbm{1} \otimes \ket{+}\bra{+}) u_{\textrm{H}} = \, \vcenter{\hbox{\includegraphics[width=0.065\linewidth]{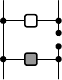}}} \,=\, \vcenter{\hbox{\includegraphics[width=0.065\linewidth]{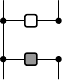}}} \,=\, \vcenter{\hbox{\includegraphics[width=0.065\linewidth]{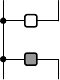}}}\,.
\end{align}
The horizontal gate exactly cancels a horizontal gate in the full lattice, introducing a corner from which the state $\ket{+}$ can propagate using the derivation from \eqref{eq:Rmatrix_plus}. The derivation for the left-moving glides is purely analogous. 

From the above derivation we can directly construct sets of conserved charges as
\begin{align}
Q_{k,+}^{(n_1\dots n_{k-1})} = \sum_{j} g^{(+)}_{j} g^{(n_1)}_{j+1} \dots g^{(n_{k-1})}_{j+k-1} g^{(+)}_{j+k}, \qquad n_l \in \{0,+\}, \\
Q_{k,-}^{(n_1\dots n_{k-1})} = \sum_{j} g^{(-)}_{j} g^{(n_1)}_{j+1} \dots g^{(n_{k-1})}_{j+k-1} g^{(-)}_{j+k},   \qquad n_l \in \{0,-\},
\end{align}
where we have introduced the notation $g^{(0)}_l = \mathbbm{1}$ for convenience and $Q_{k,\pm}^{(n_1\dots n_{k-1})}$ is the sum of local terms acting on $k+2$ sites. These operators consist of sums of local operators as products of solitons, for each of which the dynamics acts as a simple translation, with the outer edges of the operators constrained to be nontrivial in order to have full support on $k+2$ sites.

While these operators form an extensive set, they are not guaranteed to be exhaustive. For the circuit of Fourier matrices  with $u_\text{H} = F^{\dagger}$ and $u_\text{V} = F$ a set of right (left) moving gliders were obtained in \eqref{eq:two-site} as $Z_j^n X_{j+1}^{-n}$ ($X_{j}^{-n}Z_{j+1}^n $) with $n=0,1 \dots ,q-1$, since powers of gliders are again gliders.
These operators have a similar structure as the gliders~\eqref{eq:Had_gliders}, with e.g. the right moving glider acting as a diagonal matrix on site $j$ and as a nondiagonal matrix on site $j+1$. However, these are not identical. Rather, for this choice of Hadamard matrix we have that the gliders obtained in this section can be written as a linear combination of the already obtained gliders,
\begin{align}
g_{j}^{(+)} = \sum_{n=0}^{d-1} Z_j^n X_{j+1}^{-n}, \qquad g_j^{(-)} = \sum_{n=0}^{d-1}Z_{j+1}^n X_{j}^{-n}\,.
\end{align}
These identities can be directly checked by comparing the matrix elements of both sides, e.g. for the right moving glider
\begin{align}
\vcenter{\hbox{\includegraphics[width=0.08\linewidth]{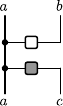}}} = F_{ab} F^{\dagger}_{ac}= \omega^{a(b-c)},\qquad \sum_{n=0}^{d-1}  Z^n_{aa} X_{bc}^{-n} = \sum_{n=0}^{d-1} \omega^{an} \delta_{b-n = c \, \textrm{mod}\,d} \, =\omega^{a(b-c)}\,.
\end{align}
No additional gliders can however be constructed from powers of $g_j^{(\pm)}$, since these are projectors. This is readily apparent from their definition as unitary transformation of e.g. the projector $\mathbbm{1} \otimes \ket{+}\bra{+}$, or can be graphically checked as
\begin{align}
\left[g_j^{(+)}\right]^2 = \vcenter{\hbox{\includegraphics[width=0.06\linewidth]{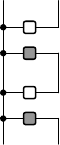}}} \propto
\vcenter{\hbox{\includegraphics[width=0.06\linewidth]{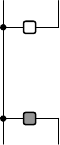}}} = g_j^{(+)}\,.
\end{align}

This result illustrates an important point: The underlying Hadamard structure directly implies the existence of a right and left moving glider of the form~\eqref{eq:Had_gliders}, but specific choices of Hadamard matrices can give rise to additional gliders and resulting conservation laws, as is the case for Fourier matrices. In the latter case we have established that the set of gliders is complete and the dynamics is hence integrable as well as Clifford, whereas an incomplete set of gliders would rather result in Hilbert space fragmentation.

\subsection{Yang-Baxter relation} \label{sec:yb}

The previous section established an extensive set of conserved charges for the full class of circuits for any choice of symmetric Hadamard matrix. Furthermore, in \sref{sec:no-cat} we constructed a complete set of gliders and hence established integrability whenever these Hadamard matrices are equivalent to Fourier matrices. The difference between completeness and incompleteness of the gliders can also be observed on the level of the unitary gates, where integrability is typically implied by an underlying Yang-Baxter or braiding equation. 
A unitary unit cell $U$ for the Hadamard circuit can be identified as
\begin{equation}\label{eq:Rmatrix}
\vcenter{\hbox{\includegraphics[width=0.3\linewidth]{fig_Rmatrix}}}\,\,,
\end{equation}
which is also the local gate in the corresponding brickwork circuit. 
For the integrable kicked Ising model at the self-dual point and the kicked Potts model for $d=3$, this gate indeed satisfies the set-theoretical Yang-Baxter relation $U_{12}U_{23}U_{12} = U_{23} U_{12} U_{23}$, graphically represented as
\begin{equation}\label{eq:YBE}
\vcenter{\hbox{\includegraphics[width=0.24\linewidth]{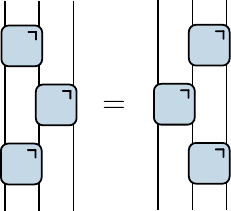}}}\,.
\end{equation}
These results track with the literature on integrable dual-unitary models~\cite{bertini2019exact,claeys2021ergodic,borsi_construction_2022,gombor_superintegrable_2022,giudice_temporal_2022}: for all known models the unitary gate satisfies the set-theoretic Yang-Baxter equation. This is again a more restricted form of integrability. In order to be integrable the gate only needs to satisfy the parameter-dependent Yang-Baxter equation~\cite{vanicat_integrable_2018}, of which the set-theoretic Yang-Baxter equation is a particular limit. Ref.~\cite{gombor_superintegrable_2022} discussed the dynamics of dual-unitary models whenever the underlying gate satisfies \eqref{eq:YBE} and is in involution, i.e. $U_{12}^2 = \mathbbm{1}$, a property we do not require here, and showed how they gave rise to superintegrable dynamics.

Remarkably, unitary gates of the form~\eqref{eq:Rmatrix} constructed from a symmetric complex Hadamard matrix satisfy the set-theoretical Yang-Baxter relation~\eqref{eq:YBE} whenever $q<6$ but not necessearily for $q \geq 6$. 
That this result breaks down at $q=6$ follows from the fact that there exist exhaustive characterizations of complex Hadamard matrices for $q <6$ but not for $q \geq 6$.  The two-site gates~\eqref{eq:Rmatrix} satisfy the set-theoretical Yang-Baxter equation whenever the underlying Hadamard matrix is given by a (possibly dephased) Fourier matrix. Furthermore, the Yang-Baxter equation is also satisfied when the building block is given by a $q=4$ complex Hadamard matrix $F_4^{(1)}(a)$ [Eq.~\eqref{eq:F4_a}], resulting in a one-parameter family of complex unitary matrices satisfying \eqref{eq:YBE}. Both can be verified by direct calculation. Crucially, for $q=2,3,5$ all complex Hadamard matrices are equivalent to Fourier matrices, and for $q=4$ these are either equivalent to Fourier matrices or to Eq.~\eqref{eq:F4_a}, implying that for $q < 6$ all gates of the form~\eqref{eq:Rmatrix} will satisfy the braiding relation. For $q=6$ it is however possible to find counterexamples, such that this is not a general property.

Symmetric complex Hadamard matrices for arbitrary $q$ can be numerically obtained in a relatively straightforward way by direct extension of the Sinkhorn algorithm from~\cite{rather_creating_2020,rather_construction_2022}. Starting from a random matrix, it is possible to iteratively generate a symmetrix complex Hadamard matrix by (i) taking the polar decomposition of the matrix, (ii) symmetrizing the unitary matrix obtained from this decomposition, and (iii) dividing each matrix element in this matrix by its absolute value and (iv) repeating these steps until convergence. If this procedure converges, which generically happens, the resulting matrix will be a symmetric complex Hadamard matrix.

As an additional verification, for $q<6$ all symmetrix complex Hadamard matrices generated in this way satisfy the Yang-Baxter equation~\eqref{eq:YBE}, whereas for $q \geq 6$ no such matrices generated from a random seed were found to satisfy the Yang-Baxter equation~\eqref{eq:YBE}.
A random example of a symmetric complex Hadamard matrix obtained in this way, which does not give rise to a unitary gate satisfying the set-theoretical Yang-Baxter equation, is given by
\begin{align}
\resizebox{.9 \textwidth}{!} 
{$\left(\begin{array}{*{6}c}
0.894+0.449i & -0.311-0.951i & 0.616-0.788i & 0.746+0.665i & 0.675+0.738i & 0.138+0.99i \\
-0.311-0.951i & 0.455-0.891i & -0.068+0.998i & -0.991-0.132i & 0.963+0.269i & 0.699+0.716i \\
0.616-0.788i & -0.068+0.998i & 0.533+0.846i & 0.458+0.889i & -0.909+0.417i & 0.949+0.314i \\
0.746+0.665i & -0.991-0.132i & 0.458+0.889i & -0.348-0.937i & 0.18+0.984i & 0.197-0.98i \\
0.675+0.738i & 0.963+0.269i & -0.909+0.417i & 0.18+0.984i & 0.991-0.131i & 0.4-0.916i \\
0.138+0.99i & 0.699+0.716i & 0.949+0.314i & 0.197-0.98i & 0.4-0.916i & 0.55+0.835i \\
\end{array}\right)$
}
\end{align}
The corresponding brickwork circuit can be numerically checked to only support gliders of the form~\eqref{eq:Rmatrix_plus}, as opposed to the $q$ right and left moving gliders obtained for Fourier matrices.

\section{Conclusions} \label{sec:conclude}

In this work we have studied a class of space-time dual models defined in terms of Hadamard matrices. These models can be viewed through the lens of recent activity on dual-unitarity, in both the brickwork and the interactions round-a-face language, as well as integrable stroboscopic quantum dynamics. Choosing the Hadamard matrices to be dephased Fourier matrices returns models that are closely connected to older work on (Clifford) cellular automata and classical spatiotemporal chaos, while also admitting integrable points.

There are various connections that are natural to explore further. The choice of complex Hadamard matrices allows for additional graphical identities as compared to the usual dual-unitarity, similar to the ZX calculus -- where delta tensors and Hadamard matrices for $q=2$ are among the basic building blocks. It would be interesting to relate the graphical ZX calculus to the one for more general complex Hadamard matrices and explore the Clifford limit, where the graphical ZX rules are known to be complete.
It would additionally be interesting to explore the dynamics for specific non-Fourier complex Hadamard matrices, e.g. multi-unitary ones. 
Different families of classical cellular automata can be obtained by constructing an inhomogeneous lattice out of (possibly dephased) Fourier matrices and their hermitian conjugate, with the resulting cellular automata satisfying dual-injectivity as a classical equivalent of dual-unitarity, and it would be interesting to explore the classical consequences of dual-injectivity.
These models can also be directly extended to other lattice geometries and higher dimensions, which will be the subject of upcoming work.

\section*{Acknowledgments}
 P.W.C. gratefully acknowledges useful discussions with M.A. Rampp and S.A. Rather. A.L. was supported by EPSRC Critical Mass Grant EP/V062654/1.

\bibliographystyle{MyBibTexStyle}
{\footnotesize
\bibliography{space_time_cats}
}
\end{document}